\documentstyle[prb,aps,epsf,twocolumn]{revtex}
\begin{document}
\draft

\title{Novel structures of Co-Cu bimetallic clusters and their magnetic properties}

\author{Jinlan Wang$^{1,2}$, Guanghou Wang$^1$, Xiaoshuang Chen$^2$, Wei Lu$^2$, Jijun Zhao$^3$}

\address{$^1${\it National laboratory of solid state microstructures and department of physics, Nanjing University, China}\\
$^2${\it National laboratory for infrared physics, Shanghai institute of technical physics, Chinese academy of sciences, China}\\
$^3${\it Department of Physics and Astronomy, University of North Carolina at Chapel Hill, Chapel Hill, NC 27599-3255}}
\maketitle

\begin{abstract}
The structural and magnetic properties of Co$_{18-m}$Cu$_m$ ($0\leq m\leq 18$%
) clusters are investigated with a genetic algorithm and a $spd$-band model
Hamiltonian in the unrestricted Hartree-Hock approximation respectively. In
general, Cu atoms tend to occupy the surface, while Co atoms prefer to the
interior of the clusters. Layered structures appear in some clusters with
given stoichiometric compositions. The introduction of Cu atoms leads to
large increase of the magnetic moment of Co-rich circumstance and nearly
zero magnetism of the Cu-rich ambient. The interaction between Cu and Co
atoms induces nonzero magnetic moment for Cu atoms. The total magnetic
moments tend to decrease with the increase of Cu atoms. However, some
particular large magnetic moment are found to be closely related to the
structures. The environment of Cu and Co atoms have a dominant effect on the
magnetism of the cluster.

\end{abstract}


Bimetallic clusters are an exciting research field due to their potential
applications in the automobile industry and oil refined as catalysts\cite
{1,2}. Such nanoscale alloys may present a number of structures and phases
that are different from those of corresponding pure metals. Previously,
there are intensive studies on homogeneous metallic clusters. But the
reports on the bimetallic clusters are scarce \cite{3,4,5,6,7,8,9},
especially for transition-metal bimetallic clusters because of the
complexity in their electronic structure.

In this report, we exploit the structural and magnetic properties of
bimetallic Co-Cu clusters. The main reason for choosing Co-Cu is that the
physical properties of bulk Co and Cu are very different. We may get a clear
picture of the various properties of the bimetallic clusters versus the
different composition ratios. Moreover, Co and Cu alloys are non-miscible.
The clusters may give a qualitative analysis from the mecoscopical points.

Although the reliable results on clusters can be obtained on basis of
quantum chemistry or density function theory\cite{6,7,8,9}, the well-known
NP problem leads to expensive computational costs. Alternatively, empirical
potential fitted from the bulk materials have been extensively employed to
study the structures and properties of clusters\cite{4,5,10,11,12,13,14}. In
this letter, we obtain the lowest energy structures of Co$_{18-m}$Cu$_m$ ($%
0\leq m\leq 18$) by a genetic algorithm (GA) with a Gupta-like many-body
potential\cite{11}. The parameters for inhomogeneous Cu-Co interaction are
derived from the average of the Cu-Cu and Co-Co parameters. In the GA scheme%
\cite{15,16,17,18}, a number of random initial configurations are generated
in the beginning. Then any two candidates in the population can be chosen as
parents to generate a child cluster by mating operation. The obtained child
cluster can be selected to replace its parent, if it has lower binding
energy but its configuration is different from any one in the population.

We have checked the validity of current parameterization by {\em ab initio}
calculation on the smallest clusters, i.e., homogeneous and inhomogeneous
dimers and trimers. The {\em ab initio} calculation is performed by using
DMol package based on density functional theory (DFT) \cite{19}. During the
DMol electronic structure calculations, the effective core potential (ECP)
and a double numerical basis including $d$-polarization function (DND) are
chosen. The density function is treated within the generalized gradient
approximation (GGA) \cite{20} with exchange-correlation potential
parameterized by Wang and Perdew\cite{21}. A direct comparison of the {\em 
ab initio} and empirical results on the structural information such as
equilibrium bond length and bond angle for those small clusters is given in
Table. I. One can find that all the bond angles of either homogeneous or
inhomogeneous trimers are well described by empirical potential. Except Co$%
_2 $ dimer, the difference of bond length between DFT and empirical
calculation is less than $0.05$\AA . We have further verified our empirical
scheme by calculating the clusters Co$_{13}$ and Cu$_{13}$. The average bond
length and average binding energy per atom of Co are $2.45$\AA\ and $3.22$
eV, which are $2.44$ \AA\ \cite{22} and $3.66\pm 0.36$ eV \cite{9} from
first-principles calculations. For Cu$_{13}$, the bond length and the
binding energy are $2.50$\AA\ and $2.59$ eV , in agreement with the results
of TB-LMTO, $2.52$\AA\ and $2.46$ eV \cite{23}. From the above comparisons,
the overall agreement of our present model potential with accurate {\em ab
initio} is rather reasonable. Therefore, we can use such Gupta-like
potential in the global structural optimization of 18-atom Cu-Co bimetallic
clusters, in which {\em ab initio} calculations up to long time scale is
computational prohibitive.

\begin{table}[tbp]
Table I. Bond length (d) and bond angle ($\theta $) are compared with the
spin-polarized DFT-GGA method (in parenthesis) for small Co, Cu or Co-Cu
clusters.
\par
\begin{center}
\begin{tabular}{cccccc}
& Co$_2$ & Cu$_2$ & CuCo &  &  \\ \hline
d (\AA) & 2.203 (2.107) & 2.235 (2.225) & 2.219 (2.258) &  &  \\ \hline
& Co$_3$ & Cu$_3$ & Co$_2$Cu & Cu$_2$Co &  \\ \hline
d (\AA) & 2.230 (2.262) & 2.337 (2.348) & 2.308 (2.361) & 2.324 (2.388)
&  \\ 
$\theta $ & 60.0 (60.0) & 60.0 (60.0) & 61.4 (62.1) & 58.8 (56.5) & 
\end{tabular}
\end{center}
\end{table}

Fig.1 gives the morphology structures corresponding to different
stoichiometric composition of Co$_{18-m}$($0\leq m\leq 18)$. Great
modifications are found in the bimetallic clusters. The most stable
structures for the clusters with $m=0,2,15-18$ are double icosahedron minus
an atom in the layer, while the rest ones prefer to the bell-like
structures. These imply that the mixing process has a great influence on the
ground state structures. This may be original from the fact that $18$-atom
constitutes a double icosahedron with a defect. The existence of a defect
may easily induce a structural transition. Hence, we can alter the
composition ratios to attain some new structures.

\begin{figure}
\vspace{-0.15in}
\centerline{
\epsfxsize=3.5in \epsfbox{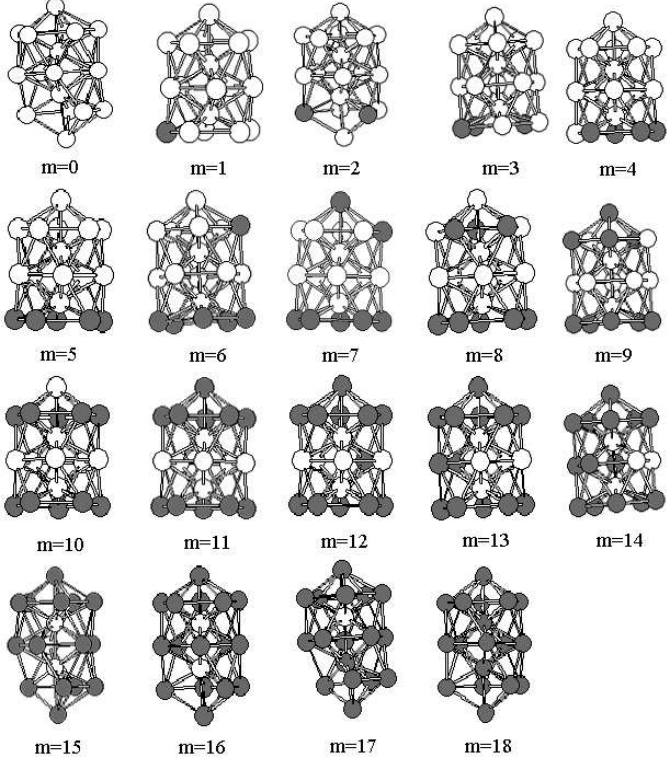}
}
\caption{Optimized structures of Co$_{18-m}$Cu$_m$ with various concentrations.}
\end{figure}

Segregation effect is found in the ground state structures of bimetallic
clusters: Cu atoms tend to occupy the surface, while Co atoms prefer to the
interior. The early occupied sites by Cu atoms have lower coordination
number (CN), then the higher CN. Take Co$_5$Cu$_{13}$ as an example, eleven
Cu atoms occupy the sites with CNs being 5 or $6$, the rest 3 Cu atoms are
in the layer with CNs=$8$. Moreover, the same kinds of atoms tend to
assemble together in the same layer. The assembling Cu atoms tend to
maximize the number of Cu-Cu and Co-Co bonds. This phenomenon may be due to
the great difference of the surface energy and the cohesive energy between
Co and Cu. To minimize the total energy, the atom with the smaller surface
energy and cohesive energy tends to occupy the surface, while the atom with
a higher surface energy and cohesive energy favors to the interior. The
average cohesive energy and surface energy of the bulk Cu, $3.544$ eV and $%
1.934$ Jm$^{-2}$, are smaller than those of the bulk Co, $4.386$ eV and $%
2.709$ Jm$^{-2}$. Another possible reason is the atomic size effect. In our
simulations, the first-nearest distance of Cu is $2.556$\AA\ , larger than
that of Co $2.507$\AA . Thus, Co are more easily surrounded by Cu atoms.

Another interesting finding is the appearance of layered structures in the
bimetallic clusters. For Co$_{13}$Cu$_5$, five Cu atoms occupy the lower CN
sites while thirteen Co atoms constitute an icosahedron. Similar features
are found in the clusters Co$_8$Cu$_{10}$ and Co$_7$Cu$_{11}$,etc. These
imply the existence of some ordering effects to maximize the number of Co-Cu
bond. The ordering effect and the segregation effect coexist and compete
with each other in the cluster, thus lead to the appearance of layered
structure and the segregation of Cu atoms. The bizarre structure characters
may bring some peculiar properties. Our previous studies have shown some
peculiar thermal behavior in Cu-Co bimetallic clusters\cite{24}. In the
following, we investigate the electronic and magnetic properties by
parameterized unrestricted Hartee-Fock approximation.

The Hamiltonian, written in a local orbital basic set, has the expression:

\begin{equation}
H=\sum\limits_{i,\alpha ,\sigma }\epsilon _{i\alpha \sigma }\stackrel{\wedge 
}{n}_{i\alpha \sigma }+\sum\limits_{\stackrel{i\not{=}j}{\alpha ,\beta
,\sigma }}t_{ij}^{\alpha \beta }\stackrel{\wedge }{c}_{i\alpha \sigma }^{+}%
\stackrel{\wedge }{c}_{i\beta \sigma }
\end{equation}
where $\stackrel{\wedge }{c}_{i\alpha \sigma }^{+}$( $\stackrel{\wedge }{c}%
_{i\beta \sigma }$) are the creation (annihilation) operators and $\stackrel{%
\wedge }{n}_{i\alpha \sigma }$is the number operator of an electron.. The $%
t_{ij}^{\alpha \beta }$is the hopping integral between different sites and
orbitals. The orbital state $\alpha $ involved in calculation includes $%
s,p_x,p_y,p_z,d_{xy}$,$d_{yz}$,$d_{xz}$,$d_{x^2-y^2}$,$d_{3z^2-r^2}$. The
single-site energy $\epsilon _{i\alpha \sigma }$ is given by 
\begin{equation}
\epsilon _{i\alpha \sigma }=\epsilon _d^0+U\Delta n(i)-\frac 12\sigma J\mu
(i)+\sum\limits_{j\not{=}i}\Delta n(j)V_{ij}
\end{equation}
Here $\epsilon _d^0$ refers to the orbital energy levels in the paramagnetic
solutions of the bulk. $\Delta n(j)$ denotes the charge change. The Coulomb
interaction $V_{ij}$ is described as

\begin{equation}
V_{ij}=\frac U{1+(UR_{ij}/e^2)}
\end{equation}
The orbital energy and the hopping integrals are taken to be the bulk values
obtained from Andersen's linear muffin-tin orbital atomic sphere
approximation(LMTO-ASA) paramagnetic bands\cite{Andersen}. The hopping are
assumed to be spin independent and are averaged for the heteronuclear.
Exchange integrals other than $J_{dd}$ are neglected and $J_{dd}(Co)=0.99$eV%
\cite{jdd}. The direct integral $U_{dd}^{}(Co)$ is obtained from Ref.\cite
{udd} , and $U_{ss}^{}/U_{dd}^{}$ relations are from the atomic tables. We
take $U_{dss}^{}=U_{pp}^{}=U_{sp}^{}$ and $U_{sd}^{}=U_{pd}^{}=\frac{%
U_{ss}^{}+U_{dd}^{}}2$\cite{fab}. For Cu, all the parameters come from Ref.%
\cite{fab}.

The magnetic moment can be determined by integrating the majority and
minority local densities of state(LDOS) up to Fermi energy: 
\begin{equation}
\mu _{i\alpha }=\int_{-\infty }^{\epsilon _F}[\rho _{i\alpha _{\uparrow
}}(\epsilon )-\rho _{i\alpha _{\downarrow }}(\epsilon )]d\epsilon \text{,}
\end{equation}
The LDOS is directly related to the diagonal elements of the local Green
function by means of the recursion method\cite{recursion}: 
\begin{equation}
\rho _{i\alpha \sigma }=-\frac 1\pi 
%TCIMACRO{\func{Im} }
%BeginExpansion
\mathop{\rm Im}%
%EndExpansion
[G_{i\alpha \sigma ,i\alpha \sigma }(\epsilon )]\text{.}
\end{equation}

Fig.2 gives the total magnetic moment of Co$_{18-m}$Cu$_m$ bimetallic
clusters as a function of the concentration of Cu atoms. The total magnetic
moments decrease from 33.61$\mu _b$ to 0$\mu _b$ as the Cu concentrations
vary from $m=2$ to $m=13$. The curve can be divided to three sections. The
first section is the Co-rich circumstance with $m=1-5$, where the
introduction of a small amount of Cu atoms enhances the magnetism of the
clusters. The contribution to the magnetism mainly comes from the Co atoms
far away from Cu atoms. It may be due to the large charge transfer between
Cu and Co atoms. The second section is the comparative composition ratio
with $m=6-12$, in which the magnetic moments fluctuate with the cluster
size. For the clusters with concentrations $m=6-8$, the case is very similar
to the first. But in the case of $m=9-11$, the interior pentagonal bipyramid
significantly contributes to the magnetism of the cluster. Thus, relatively
large magnetic moments are found for these clusters. The third section is
the Cu-rich clusters with $m=13-17$. Cu atoms seem to have a ''screen''
effect on the magnetic moments and lead to zero magnetic moment in these
clusters.

\begin{figure}
\vspace{0.65in}
\centerline{
\epsfxsize=3.0in \epsfbox{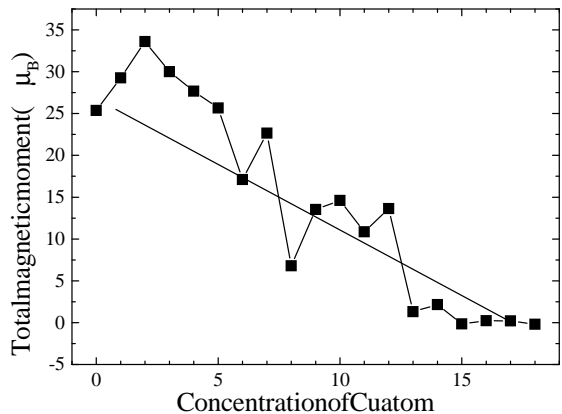}
}
\vspace{-0.75in}
\caption{The total magnetic moments of Co$_{18-m}$Cu$_m$ bimetallic clusters
as a function of the concentrations of Cu atoms.}
\end{figure}

The magnetism of the cluster is found closely related to the environments of
Co and Cu atoms and the cluster geometrical characters. For example, the
total magnetic moment of Co$_{16}$Cu$_2$ is much larger than those of Co$%
_{17}$Cu$_1$ and Co$_{15}$Cu$_3$. The main reason is due to the large charge
transfer between Co-Cu. Obviously, the number of Co-Cu bond in Co$_{16}$Cu$%
_2 $ is much more than that of Co$_{17}$Cu$_1$ and Co$_{15}$Cu$_3$. The more
the number of Co-Cu bond in the cluster, the larger the charge transfer,
which induces a large magnetic moment. Similarly, the large magnetic moments
for Co$_{11}$Cu$_7$ and Co$_8$Cu$_{10}$ are obtained, compared with Co$_{10}$%
Cu$_8$ and Co$_7$Cu$_{11}$. Contrary, the total magnetic moments for Co$%
_{13} $Cu$_5$ is found much larger than that for Co$_{12}$Cu$_6$, even the
number of Co-Cu bond in the former is more than the latter one. As discussed
above, Co$_{13}$Cu$_5$ is a good layered structure with 5 Cu atoms in the
same layer and 13 Co atoms constituting an icosahedron. Although the high
symmetry compresses the magnetism of the clusters, the 5 Cu atoms has an
less effect to the icosahedron. It is well-known that the smaller the
cluster size, the larger the magnetic moments. Therefore, a large magnetic
moment is obtained for the cluster Co$_{13}$Cu$_5$.

\begin{figure}
\vspace{0.65in}
\centerline{
\epsfxsize=3.0in \epsfbox{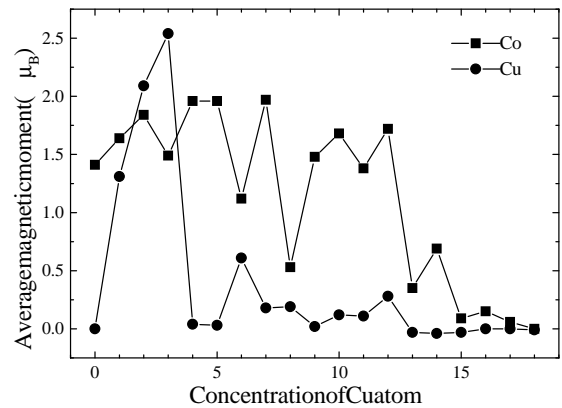}
}
\vspace{-0.75in}
\caption{Average magnetic moments of Co and Cu atoms as a function of the concentrations of Cu atoms.}
\end{figure}

It is also worthy to note the nonzero magnetic moment of Cu atoms in some
bimetallic clusters. Fig.3 shows the average moments of Co and Cu as a
function of the concentration of Cu. Although the average magnetic moment of
Cu atoms is zero or nearly zero in most cases, particularly high magnetic
moments are found in the clusters with $m=1-3$ and less pronounced peaks are
found at $m=6,12$. For the case of Co atoms, the hybridization with the Cu
atoms leads to an oscillatory behavior for the average magnetic moment.
These may be due to the different charge transfer. For Co atoms, the charges
are transferred from $sp$ orbitals to $d$ orbital, contrary to the case of
Cu atoms that the charges are transferred from $d$ orbital to $sp$ orbitals.
Further, the charge transfer takes place from Co atoms to Cu atoms for the
clusters with Cu concentration $m<13$, while the case is reverse for $%
m\geq 13$. For the case of $m=3,6,12$, the charge transfer from Co
atoms to Cu atoms is found very large, which induces the large magnetism of
Cu atom.

To explore the origin of peculiar magnetic properties, we show the total
density of states(DOS) and $sp$,$d$ DOS of the pure and represented
bimetallic clusters in Fig.4. The cluster Fermi level is presented as a
dashed vertical line and shifted to zero. In general, the DOS near to Fermi
level play a primary role in determining the magnetism of the clusters.
Obviously, the contribution of $d$ electrons is found dominantly, while the $%
sp$ electrons contribution is low. The contribution of $d$ electrons in Co$%
_{16}$Cu$_2$ is larger than that in Co$_{18}$, which leads to a large
increase of the magnetic moments in Co$_{16}$Cu$_2$. Similarly, the
contribution of $d$ electrons in Co$_{16}$Cu$_2$ is also larger than that in
Co$_{13}$Cu$_5$, thus their corresponding magnetism are different from each
other. For Co$_{10}$Cu$_8$, the contribution of $d$ electrons is relatively
less near to the Fermi level compared with other cases, which leads to a
particular small magnetic moments. Moreover, the hybridization between $sp$
and $d$ orbitals among Co-Cu, Co-Co and Cu-Cu atoms in Co$_{10}$Cu$_8$ is
also smallest among these four cases, while it is largest in Co$_{16}$Cu$_2$%
. This also enhances the magnetism of Co$_{16}$Cu$_2$.

\begin{figure}
\vspace{0.65in}
\centerline{
\epsfxsize=3.5in \epsfbox{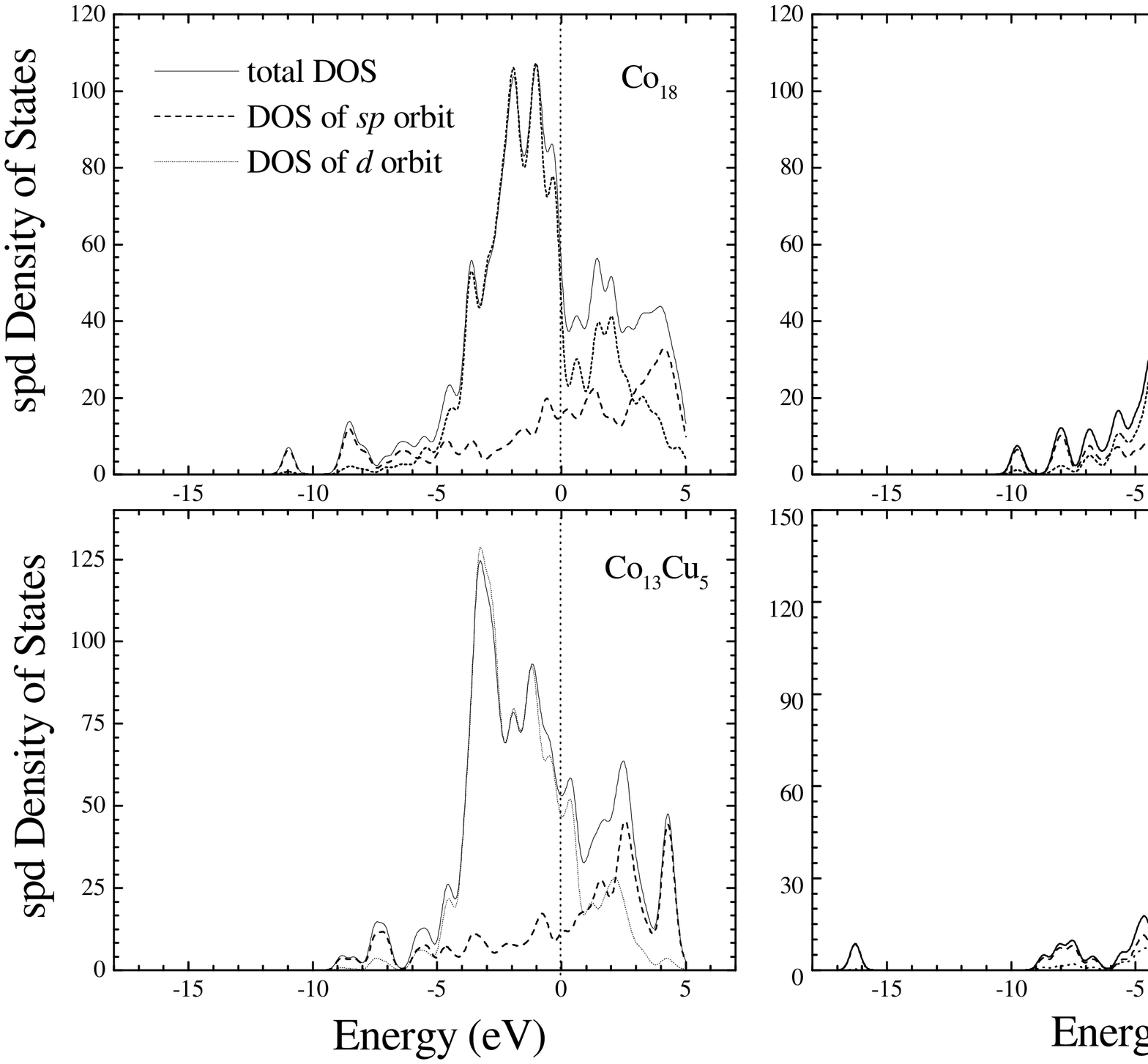}
}
\vspace{-0.85in}
\caption{Total,sp and d orbitals DOS for some compositions: (a) Co$_{18}$;
(b) Co$_16$Cu$_{2}$;(c) Co$_{13}$Cu$_5$, (d)Co$_{10}$Cu$_{8}$; The vertical
dashed lines indicate the DOS integrated Fermi level.}
\end{figure}

In summary, the geometrical and magnetic properties of bimetallic clusters Co%
$_{18-m}$Cu$_m$ have been studied by a genetic algorithm and a spin
polarized tight-binding Hamiltonian. The main conclusions can be made as
follows: (1) Great modifications are found for the $18$-atom bimetallic
clusters due to different composition ratios. This suggests that we can
alter the composition ratios to attain new structures. (2) Ordering effect
and segregation effect influence the atomic configurations of the bimetallic
clusters simultaneously, which leads to the segregation of the Cu atoms to
the surface and some layered structures. These may explain why there are no
corresponding ordered compounds of Co-Cu bulk in the low temperature. (3)
The introduction of Cu atoms causes a dramatic increase of magnetism in
Co-rich circumstance and nonzero moments for Cu atoms. Particular large
moments can be associated with the environments of Co and Cu atoms and the
geometrical characters. (4) Cu atoms have an ''screen'' effect on the
cluster magnetism in Cu-rich ambient and enhance the magnetics in Co-rich
environment.

This work is financially supported by the National Natural Science
Foundation of China (No.29890210, 100230017).

\end{document}